\documentclass{PoS}

\title{Hunting exotics}

\ShortTitle{Hunting exotics}

\author{\speaker{Teresa F. Caram\'es}\\
        Universidad de Salamanca\\
        E-mail: \email{carames@usal.es}}

\author{Alfredo Valcarce\\
        Universidad de Salamanca\\
        E-mail: \email{valcarce@usal.es}}

\author{Javier Vijande\\
        Universidad de Valencia\\
        E-mail: \email{javier.vijande@uv.es}}

\abstract{In this work we study the $QQ\bar{n}\bar{n}$ systems from both four--quark and meson--meson molecule formalisms, arriving to consistent conclusions when the proper bases are chosen. These states are interesting as they are thought to be found in experiments soon at LHC or RHIC. Such a source of experimental information will significantly improve our current understanding of charm spectroscopy. 
}

\FullConference{Sixth International Conference on Quarks and Nuclear Physics,\\
		April 16-20, 2012\\
		Ecole Polytechnique, Palaiseau, Paris}

\begin{document}
After the discovery of $J/\Psi$ in 1974 \cite{Aubert:1974js,Augustin:1974xw}, three quiet decades followed in charmonium spectroscopy. Although many states belonging to the $\bar{c}c$ spectrum were discovered, all of them nicely fitted into the slots theoretically predicted. The situation was analogous for $D$ and $D_s$ mesons: their spectra filled up without difficulties. However, two new discoveries performed in 2003 meant a turning point. The first one, a charm--strange state $D_{sJ}^*$(2317), was reported by BABAR \cite{Aubert:2003fg} and promptly confirmed by different collaborations. Its small width was not compatible for such a small mass if assumed a standard $\bar{q} q$ configuration. The second one is the celebrated X(3872), first reported by BELLE \cite{Choi:2003ue}. Although both states are firmly stablished, they can hardly be understood in a simple $q \bar q$ scheme, what makes their nature not being definitively settled.

By having a look at the charmonium spectrum, one can check that things look easy below the threshold for the $\bar{D} D$ production: all the states match the predictions for the $\bar{c} c$ spectrum based on a simple color Fermi-Breit interaction. Above this threshold, it gets all more involved, with many puzzles awaiting solution. There are still empty levels of the $\bar{c} c$ spectrum for whom a candidate has not been found, whereas several of the abundant new states do not fit in the $\bar{c} c$ spectrum, as the confirmed X(3872) or Y(4260), or the still dubious Z(4050) or Z(4250). These latter are of particular interest since they have been interpreted as charged resonances and therefore cannot belong to the charmonium spectrum. Thus we have a zoo of particles named X, Y or Z whose structure has yet to be determined. Some possible explanations have been postulated so far: molecular, $\bar{c} c$ + molecular, hybrid, or $\bar{c} c$ + higher order Fock space components, among others. The wave function of a meson ($B$ = 0) within the constituent quark model, where explicit gluon degrees of freedom are frozen in terms of a quark constituent mass, can be written as: 
\begin{equation}
|\Psi (B=0) \rangle \, = \, \alpha_1 \, |q \bar{q} \rangle \, + \, \alpha_2 \, |qq\bar{q}\bar{q} \rangle \, + \, ...
\end{equation}
where $\sum_i |\alpha_i|^2 =1$. Ground state $\bar{q}q$ mesons have negative parity. Positive parity mesons can be reached either through a four--quark configuration in a $L = 0$ state or by adding a unit of orbital angular momentum to the $\bar{q} q$ pair. In the constituent quark model, the mass of such a pair is around 600 MeV \cite{Diakonov:1985eg}, whereas the mass shift produced by a unit of $L$ stands around 500--600 MeV \cite{Nakamura:2010zzi}. None of these two mechanisms is suppressed by the other and therefore the four quark piece is not negligible now. The challenging consequence of this so-called unquenching of the quark model \cite{Close:2007ny}, that seems to be unavoidable nowadays when pursuing a description of the excited hadronic spectrum, is the appearance of exotic states. They could be stable in nature, with a mass that is similar to the positive parity excitations, and they could not be described by the lowest order Fock space components of the naive quark model, $|q \bar q\rangle $. This would be the case of doubly charmed or bottom mesons ( $Q Q \bar{n}\bar{n}$ states, where $Q$ stands for the heavy quark and $n$ for a light one), long ago suggested in the literature \cite{Ader:1981db}. They were found to be stable against dissociation provided the ratio between the heavy and light masses was large enough (in a clear contrast with $Q\bar{Q} n \bar{n}$, for instance). If such states were below the $D D$ threshold, they would be narrow and would clearly show up in the experiments.

Let us get more insight into these $QQ \bar{n}\bar{n}$ systems by approaching them from two different techniques. We solved in first place the Schr\"{o}dinger equation through the Hyperspherical Harmonic (HH) method \cite{Vijande:2009kj}. This is done by expanding the trial wave function in terms of HH functions, generalizing thus the well--known Spherical Harmonic formalism. The main difficulty one has to fight with is the construction of base states with the proper symmetry. The Pauli principle needs to be imposed since some of the quarks in $Q Q \bar{n}\bar{n}$ are identical. A constituent quark cluster model (CQC) has been employed for the interaction. When screening all the $J^P$ channels for the $cc \bar{n}\bar{n}$ system, only one state, the $(I) J^P = (0) 1^+$, is found to lie below the $DD$ threshold. 

To get a deeper understanding on the structure of this four--quark state, one can develop a method to compute probabilities of physical vectors in any four--body system \cite{Vijande:2009zs}. It has been found that exotic states would present significant probabilities of different physical channels. There are three different ways to couple $qq\bar{q}\bar{q}$ to a colorless state:
\begin{eqnarray}
\left[\left(q_1 q_2\right) \left(\bar q_3 \bar q_4\right)\right] &\equiv& 
  \left\lbrace | \bar 3_{12} 3_{34}\rangle, |6_{12} \bar 6_{34} \rangle\right\rbrace \equiv
  \left\lbrace | \bar 3 3 \rangle_c^{12}, |6 \bar 6 \rangle_c^{12}\right\rbrace \,, \label{b1} \\
\left[(q_1\bar q_3)(q_2\bar q_4)\right] &\equiv&
  \{|1_{13}1_{24}\rangle,|8_{13} 8_{24}\rangle\}\equiv
  \{|11\rangle_c,|88\rangle_c\} \, , \label{b2} \\
\left[(q_1\bar q_4)(q_2\bar q_3)\right]&\equiv&
  \{|1_{14}1_{23}\rangle,|8_{14} 8_{23}\rangle\}\equiv
  \{|1'1'\rangle_c,|8'8'\rangle_c\}\,. \label{b3}
\end{eqnarray}
Bases (\ref{b2}) and (\ref{b3}) contain physical vectors, i.e. the $|11\rangle$ and $|1'1'\rangle$ components, together with octet--octet (hidden color) ones. All vectors inside each basis are orthogonal, however bases are not orthogonal to each other. We aim to express the $(0)$ $1^+$ four--quark state in terms of the physical components. When solving the problem in base (\ref{b2}) we found a probability of 62.6\% to obtain a hidden--color nonphysical vector $|88\rangle_c$. However this piece can be decomposed again, as bases (\ref{b2}) and (\ref{b3}) can be connected through the following antiunitary transformation:
\begin{eqnarray}
 |11\rangle_c &=& \cos \alpha |1'1'\rangle_c + \sin \alpha |8'8'\rangle_c \, \label{eq5} \\ 
 |88\rangle_c &=& \sin \alpha |1'1'\rangle_c - \cos \alpha |8'8'\rangle_c  \,. 
\end{eqnarray}
That  would generate another physical state, $|1'1'\rangle_c$, together with a new hidden--color piece ($|8'8'\rangle_c$). This latter can be reexpanded again in terms of the first base through Eq.~(\ref{eq5}) and this process should be repeated up to infinity getting a series that can be summed up \cite{Vijande:2009zs}. After doing so one gets 50.5\% of $DD^*$ and 49.5\% of $D^*D^*$, in our present  $(I) J^P = (0) 1^+$ case, being $|11\rangle_c \equiv |DD^*\rangle$ and $|1'1'\rangle_c \equiv |D^*D^* \rangle$. One may compare the probabilities obtained for the only bound state of the $QQ\bar{n}\bar{n}$ system with those for an unbound channel, the $(I) J^P = (1) 1^+$ for instance. We find there is no probability associated to the $D^*D^*$ channel. This comparison helps us to see the relation between the binding energy and the compactness of the system. The higher the binding energy, the more compact gets a system and the more spread the probabilities over different physical channels are.
\begin{figure}[t]
\begin{center}
\includegraphics[scale=0.4]{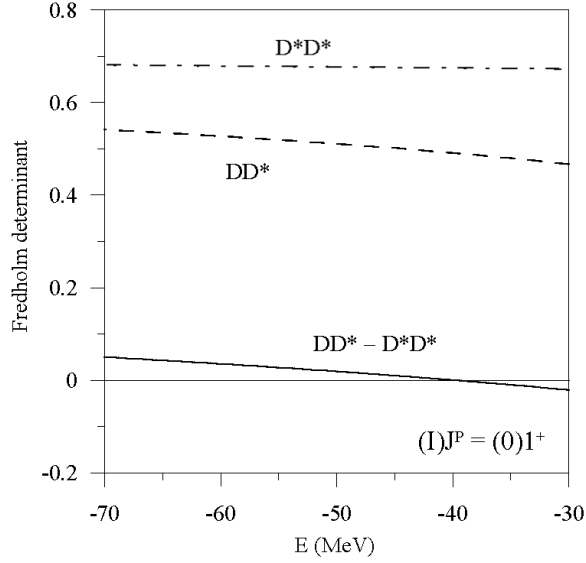}
\end{center}
\caption{Fredholm determinant for the $(I)J^{P}=(0)1^{+}$ $cc\bar n\bar n$
system. The dashed line corresponds to the calculation considering only
$DD^*$ singlet-singlet color states. The dashed--doted line stands for the results considering only the $D^*D^*$ system. The solid line represents the results
including the full set of color states or, in other words, the $DD^*-D^*D^*$ 
coupled system (see text for details).}
\label{fredholm}
\end{figure}

Let us now face the problem from the two--meson interaction point of view. For that purpose we start from a physical system made of two $D$ mesons, $M_1$ and $M_2$ ($M_i=D, D^*$), with quantum numbers $(I)J^{P}$ in a relative $S$ state. They interact through a potential $V$ that contains a tensor force. Then, in general, there is a coupling to the $M_1 M_2$ $D-$wave and to any other two $D$-meson system ($DD$, $DD^*$, $D^*D^*$) that can couple to the same quantum numbers $(I)J^{P}$. Thus, if we denote $D_1 \equiv DD$, $D_2 \equiv DD^*$ and $D_3\equiv D^*D^*$, the Lippmann-Schwinger equation for the $M_1M_2$ scattering becomes
\begin{eqnarray}
t_{\alpha\beta;ji}^{\ell_\alpha s_\alpha, \ell_\beta s_\beta}(p_\alpha,p_\beta;E)& = & 
V_{\alpha\beta;ji}^{\ell_\alpha s_\alpha, \ell_\beta s_\beta}(p_\alpha,p_\beta)+
\sum_{\begin{array}{c}{\scriptstyle \gamma=D_k} \\{\scriptstyle (k=1,2,3)}\end{array}}\sum_{\ell_\gamma=0,2} 
\int_0^\infty p_\gamma^2 dp_\gamma V_{\alpha\gamma;ji}^{\ell_\alpha s_\alpha, \ell_\gamma s_\gamma}
(p_\alpha,p_\gamma) \nonumber \\
& \times& \, G_\gamma(E;p_\gamma)
t_{\gamma\beta;ji}^{\ell_\gamma s_\gamma, \ell_\beta s_\beta}
(p_\gamma,p_\beta;E) \,\,\,\, , \, \alpha,\beta=D_1,D_2,D_3 \,\, ,
\label{eq0}
\end{eqnarray}
where $t$ is the two-body scattering amplitude, $j$, $i$, and $E$ are the angular momentum, isospin and energy of the system,
$\ell_{\alpha} s_{\alpha}$, $\ell_{\gamma} s_{\gamma}$, and $\ell_{\beta} s_{\beta }$ are the initial, intermediate, and final orbital angular momentum and spin, respectively,  and $p_\gamma$ is the relative momentum of the two-body system $\gamma$. The basic ingredient to solve the scattering problem are the interacting potentials. They are taken from the CQC model, the very same that we used for the Hyperspherical Harmonic formalism. To obtain them from the basic $\bar{q} q$ interaction we use a Born--Oppenheimer approximation. 

All positive--parity channels made by S-wave interacting mesons, up to $J^P = 2^+$, have been analyzed. Again, only one of them is attractive enough to be bound. This channel is $(I) J^P = (0) 1^+$, the same that was found to bind from our four--quark study. We can have a look at the potentials \cite{Carames:2011zz} and check that both $DD^*$ and $D^*D^*$ are attractive. It gets all more quantitative and clear when attending at the Fredholm determinant, as done in Fig.~\ref{fredholm}. Although both total potentials, $DD^*$ and $D^*D^*$ are attractive (Fredholm determinant smaller than 1), neither of them is bound. To achieve a bound system, or a Fredholm determinant equal to zero, one needs to consider the transition between $DD^*$ and $D^*D^*$, and solve the coupled--channel system. Again the coupling, as we saw from the HH formalism is the main ingredient to make the system bound. At short distances, identical quarks can recouple to different vectors of the Hilbert space. If any of them is not considered, a fundamental ingredient of the calculation would be neglected. Such an effect would never happen when dealing with hadronic degrees of freedom. In that case, we would just have an additional channel 275 MeV above, not giving any significant contribution in a coupled--channel calculation. 

The need to incorporate the complete Hilbert space is also evident when the four--quark formalism is employed: if the higher $D^*D^*$ channel was not included in the calculation, one would be artificially neglecting the large portion of vector $|88\rangle_c$ in base (\ref{b2}). If the model parameters were fitted to some observables they would necessarily include the effect of such a restricted Hilbert space and one would never be able to use them for different quantum numbers without arriving to wrong conclusions, loosing thus any predictive power. We can therefore conclude that formalisms based on four--quark and meson--meson configurations are fully compatible whenever all the relevant basis vectors are taken into account. 

When moving to the bottom sector one finds four more canditates for observation. They are $(I) J^P = (0) 1^+$, $(0)$ $0^+$, $(1)$ $3^-$ and $(0)$ $1^-$. Together with the $(0) 1^+$, the interest in these states is increasing as for the first time there are chances to observe such large mass exotic states in quite a near future: LHC may discover tetraquark states via gluon--gluon fusion due to both large number of events and their unique signature in the detectors \cite{Yuqi:2011gm}. Also at RHIC the identification of hadronic molecular states by means of relativistic heavy ion collisions has been suggested by employing the coalescence model for hadron production \cite{Cho:2010db,Ohnishi:2011nq}. If any of these suggestions gets real, the new data related to these double charm or bottom exotic systems will give a huge contribution to the heavy quark spectroscopy. 
\\

This work has been partially funded by the Spanish
Ministerio de Educacion y Ciencia and EU FEDER
under Contract No. FPA2010-21750, and by the Spanish
Consolider-Ingenio 2010 Program CPAN (CSD2007-
00042).

\end{document}